# Stability of $^{248-254}$Cf isotopes against alpha and cluster radioactivity

K. P. Santhosh* and R. K. Biju

*School of Pure and Applied Physics, Kannur University, Swami Anandatheertha Campus, Payyanur 670 327, INDIA*

**Abstract**

Stability of $^{248-254}$Cf nuclei against alpha and cluster emission is studied within our Coulomb and proximity potential model (CPPM). It is found that these nuclei are stable against light clusters (except alpha particle) and instable against heavy cluster $(A_2 \geq 40)$ emissions. For heavy cluster emissions the daughter nuclei lead to doubly magic $^{208}$Pb or neighbouring one. The effect of quadrapole and hexadecapole deformations of parent nuclei, daughter nuclei and emitted cluster on half lives are also studied. The computed alpha decay half life values (with including quadrupole deformation $\beta_2$) are in close agreement with experimental data. Inclusion of quadrupole deformation reduces the height and width of the barrier (increases the barrier penetrability) and hence half life decreases.

*email address*: drkpsanthosh@gmail.com

## 1. Introduction

Cluster radioactivity is the spontaneous emission of particle heavier than alpha particle predicted by Sandulescu et al [1] in 1980, after four years Rose and Jones [2] confirmed this phenomenon in the emission of $^{14}C$ from $^{223}Ra$ isotope. After the observation of cluster radioactivity, lots of efforts have been done on both experimental and theoretical fronts for understanding the physics of cluster radioactivity. In literature there existed old fission data of Jaffey and Hirsch [3] for $^{24}Ne$ decay of $^{232}U$, which indicates that this phenomenon was already observed in 1951, but the authors did not distinguish it from the spontaneous fission process. In a very recent experiment, Bonetti et al [4] have confirmed that the emission of $^{24}Ne$ from $^{232}U$, seen in 1951 could not be due to spontaneous fission since the then-observed cluster decay constant is larger than by an order of magnitude $10^2$ than their presently measured upper limiting value of spontaneous fission decay constant. At present about 24 modes of cluster decay from about 20 parent nuclei emitting clusters ranging from carbon to silicon are confirmed so far. For e.g. $^{14}C$ from $^{221}Fr$, $^{221-226}Ra$ and $^{226}Th$, $^{20}O$ from $^{228}Th$, $^{24,26}Ne$ from $^{230,232}Th$ and $^{232,234}U$, $^{28,30}Mg$ from $^{238}Pu$, $^{32,34}Si$ from $^{238}Pu$ and $^{241}Am$ etc. are observed.

The present study points out the role of deformations on half lives in the cluster decay process. Since the beginning of cluster radioactivity, it was recognized to be a consequence of the shell closure of one or both the fragments because of its cold nature; i.e. the low excitation energy involved in the process. An opened problem in the study of cluster radioactivity is represented by the question of the existence of only the spherical or both the spherical and deformed closed shells. It is relevant to mention here that all the parents $^{228,230}Th$, $^{232,234,238}U$, $^{236,238}Pu$ and $^{242}Cm$ and their respective emitted clusters $^{20}O$, $^{24}Ne$ and $^{28,30}Mg$ considered here are deformed, except for $^{25,26}Ne$ and $^{32,34}Si$ which are spherical or nearly spherical. Also all parent nuclei are prolate deformed whereas clusters $^{24}Ne$, $^{30}Mg$ are oblate deformed and $^{20}O$, $^{28}Mg$ are prolate deformed.

The effects of deformation in cluster decay half life are studied by many authors using different theoretical models. The theoretical study of deformation effects on the WKB penetrabilities have been carried out by Sandulescu et al [5, 6] using the double folded Michigan-3 Yukawa (M3Y) potential for a spherical daughter and a quadrupole deformed emitted cluster. In 1986 Pik Pichak [7, 8] studied the effect of ground state deformation of parent and daughter on half life treating emitted cluster as spherical. One year later Shi and Swiatecki [9] studied the effect of deformation of parent, daughter and shell attenuation on half-life time treating emitted cluster as spherical in shape. Kumar et al [10] studied the effect of deformation of cluster and daughter nuclei and also the role of neck formation in overlap region. Shanmugam et al [11] put forward cubic plus Yukawa plus exponential model (CYEM) which uses Coulomb and Yukawa plus exponential potential as interacting barrier for separated fragments and cubic potential for the overlap region. The authors also studied [12] the role of deformation of parent and daughter nuclei on half-life time.

Californium does not occur naturally but is produced artificially in nuclear reactors and particle accelerators. Californium was first produced [13-15] in 1950 in a cyclotron at the University of California at Berkeley by bombarding $^{242}$Cm with helium ions. The half-lives of californium isotopes range from 0.91s to 900 years. All the $^{248-254}$Cf isotopes decay by emitting an alpha particle but $^{248}$Cf also decay by spontaneous fission, a process in which the atom self-disintegrates into two smaller atoms accompanied by a burst of neutrons and a release of energy and nearly about 3% of the $^{252}$Cf decays by spontaneous fission. One of the interesting facts for the study of californium isotopes is that some of these isotopes show both spontaneous binary and ternary fission [16-19].

Within the Coulomb and proximity potential model (CPPM) proposed by one of us (KPS), we have studied [20] the cold valleys in the radioactive decay of $^{248-254}$Cf isotopes and the computed alpha decay half-life time values were found to be in agreement with the experimental data. In the present paper we have investigated all the possible cluster emissions from the $^{248\text{-}254}$Cf parents by including the quadrupole and hexadecapole deformations of the decaying parent nucleus along with that of emitted cluster and corresponding daughter nucleus in the ground state. The details of the model are given in section II. The Results and discussion are given in Section III and the conclusion is given in Section IV.

## 2. The Coulomb and Proximity Potential Model (CPPM)

In Coulomb and proximity potential model the potential energy barrier is taken as the sum of Coulomb potential, proximity potential and centrifugal potential for the touching configuration and for the separated fragments. For the pre-scission (overlap) region, simple power law interpolation as done by Shi and Swiatecki [21] is used. The proximity potential was first used by Shi and Swiatecki [21] in an empirical manner and has been extensively used over a decade by Gupta et al [22-25] in the preformed cluster model (PCM) which is based on pocket formula of Blocki et al [26] given as:

$$\Phi(\varepsilon) = -\left(\frac{1}{2}\right)(\varepsilon - 2.54)^2 - 0.0852(\varepsilon - 2.54)^3 \qquad \text{for } \varepsilon \le 1.2511 \qquad (1)$$

$$\Phi(\varepsilon) = -3.437 \exp\left(\frac{-\varepsilon}{0.75}\right) \qquad \text{for } \varepsilon \ge 1.2511 \qquad (2)$$

where $\Phi$ is the universal proximity potential. In the present model, another formulation of proximity potential [27] is used as given by equations (6) and (7). In this model cluster formation probability is taken as unity for all clusters irrespective of their masses, so the present model differs from PCM by a factor $P_0$, the cluster formation probability. In the present model assault frequency, ν is calculated for each parent-cluster combination which is associated with vibration energy. But Shi and Swiatecki [9] get ν empirically, unrealistic values $10^{22}$ for even A parent and $10^{20}$ for odd A parent.

We would like to mention that, the proximity potential has been revised and new proximity potentials were given with many extensions (see Ref [28-31]). But we have shown that the theoretical calculations done on the cluster half lives (see Table 2 of Ref [32]) matches very well with the experimental values and also with other models, even though the potential overestimates the barrier heights by 7-8%. This is the reason for using the 40 years old proximity potential 1977 in the present calculation.

The interacting potential barrier for a parent nucleus exhibiting exotic decay is given by

$$V = \frac{Z_1 Z_2 e^2}{r} + V_p(z) + \frac{\hbar^2 \ell(\ell+1)}{2\mu r^2} \qquad \text{, for } z > 0 \tag{3}$$

Here $Z_1$ and $Z_2$ are the atomic numbers of the daughter and emitted cluster, 'z' is the distance between the near surfaces of the fragments, $r = z + C_1 + C_2$ is the distance between fragment centers, $\ell$ represents the angular momentum, $\mu$ the reduced mass, $V_P$ is the proximity potential is given by Blocki et al [26] as

$$V_p(z) = 4\pi\gamma b\left[\frac{C_1 C_2}{(C_1 + C_2)}\right]\Phi\left(\frac{z}{b}\right) \tag{4}$$

With the nuclear surface tension coefficient,

$$\gamma = 0.9517[1 - 1.7826(N - Z)^2 / A^2] \quad \text{MeV/fm}^2 \tag{5}$$

where N, Z and A represent neutron, proton and mass number of parent. It is to be noted that new versions of the surface tension coefficient are available in Ref [30, 31]. Φ represent the universal the proximity potential [27] given as

$$\Phi(\varepsilon) = -4.41 e^{-\varepsilon/0.7176} \quad \text{, for } \varepsilon > 1.9475 \tag{6}$$

$$\Phi(\varepsilon) = -1.7817 + 0.9270\varepsilon + 0.0169\varepsilon^2 - 0.05148\varepsilon^3 \quad \text{for } 0 \le \varepsilon \le 1.9475 \tag{7}$$

With ε = z/b, where the width (diffuseness) of the nuclear surface b ≈1 and Siissmann central radii $C_i$ of fragments related to sharp radii $R_i$ is

$$C_i = R_i - \left(\frac{b^2}{R_i}\right) \tag{8}$$

For $R_i$ we use semi empirical formula in terms of mass number $A_i$ as [26]

$$R_i = 1.28 A_i^{1/3} - 0.76 + 0.8 A_i^{-1/3} \tag{9}$$

We would like to bring in to the attention of the readers that new versions of radius are available in Ref [30, 31].

For the internal part (overlap region) of the barrier a simple power law interpolation as done by Shi et al [21] is used given as,

$$V = a_0 (L - L_0)^n \qquad \text{for } z < 0 \tag{10}$$

where $L = z + 2C_1 + 2C_2$ and $L_0 = 2C$, the diameter of the parent nuclei. The constants $a_0$ and n are determined by the smooth matching of the two potentials at the touching point.

Using one dimensional WKB approximation, the barrier penetrability P is given as

$$\mathrm{P} = \exp\left\{ -\frac{2}{\hbar} \int_a^b \sqrt{2\mu (V - Q)}\, dz \right\} \tag{11}$$

Here the mass parameter is replaced by $\mu = m A_1 A_2 / A$, where m is the nucleon mass and $A_1$, $A_2$ are the mass numbers of daughter and emitted cluster respectively. The turning points "a" and "b" are determined from the equation $V(a) = V(b) = Q$. The above integral can be evaluated numerically or analytically. In the present work, numerical method has been adopted for calculating the penetrability.

The half life time is given by

$$T_{1/2} = \left(\frac{\ln 2}{\lambda}\right) = \left(\frac{\ln 2}{\upsilon P}\right) \tag{12}$$

Where, $\upsilon = \left(\frac{\omega}{2\pi}\right) = \left(\frac{2E_v}{h}\right)$ represent the number of assaults on the barrier per second and λ the decay constant. $E_v$, the empirical vibration energy is given as [33]

$$E_v = Q\left\{0.056 + 0.039\exp\left[\frac{(4-A_2)}{2.5}\right]\right\} \qquad \text{for } A_2 \geq 4 \tag{13}$$

For alpha decay, $A_2 = 4$ and therefore the empirical vibration energy becomes,

$$E_v = 0.095Q \tag{14}$$

In the classical method, the $\alpha$ particle is assumed to move back and forth in the nucleus and the usual way of determining the assault frequency is through the expression given by $\nu = velocity/(2R)$, where $R$ is the radius of the parent nuclei. But the alpha particle has wave properties; therefore a quantum mechanical treatment is more accurate. Thus, assuming that the alpha particle vibrates in a harmonic oscillator potential with a frequency $\omega$, which depends on the vibration energy $E_v$, we can identify this frequency as the assault frequency $\nu$ given in eqns. (12)-(14).

The Coulomb interaction between the two deformed and oriented nuclei taken from [34] with higher multipole deformation included [35, 36] is given as,

$$V_C = \frac{Z_1Z_2e^2}{r} + 3Z_1Z_2e^2 \sum_{\lambda,i=1,2} \frac{1}{2\lambda+1}\frac{R_{0i}^{\lambda}}{r^{\lambda+1}} Y_{\lambda}^{(0)}(\alpha_i)\left[\beta_{\lambda i} + \frac{4}{7}\beta_{\lambda i}^{2} Y_{\lambda}^{(0)}(\alpha_i)\delta_{\lambda,2}\right] \tag{15}$$

with

$$R_i(\alpha_i) = R_{0i}\left[1 + \sum_{\lambda}\beta_{\lambda i}Y_{\lambda}^{(0)}(\alpha_i)\right] \tag{16}$$

where $R_{0i} = 1.28A_i^{1/3} - 0.76 + 0.8A_i^{-1/3}$. Here $\alpha_i$ is the angle between the radius vector and symmetry axis of the i$^{th}$ nuclei (see Fig.1 of Ref [35]). Note that the quadrupole interaction term proportional to $\beta_{21}\,\beta_{22}$ is neglected because of its short range character.

## 3. Results and discussion

The calculations are done by using Coulomb potential, proximity potential and centrifugal potential for the touching configuration and for the separated fragments. It is well known that proximity potential 77 overestimates barrier heights by 7-8% [28, 29]. The Q values are computed using the experimental binding energies of Audi et al [37]. So full shell effects are contained in our model that comes from experimental mass excess. In present work the half lives are calculated for zero angular momentum transfers.

We have studied [20] the cold valleys in the radioactive decay of $^{248\text{-}254}$Cf isotopes and is found that the minima in fragmentation potential occurs at $^{4}$He, $^{10}$Be, $^{14,16}$C, $^{20,22}$O, $^{24,26}$Ne, $^{32,34}$Si, $^{40,42}$S, $^{44,46,48}$Ar, $^{48,50,52}$Ca etc. The minima in cold valley plot represent the most probable cluster emission from the corresponding parents. But from the computed half lives for these clusters we have found that these parents are stable against light clusters (except alpha particle) and instable against heavy cluster $(A_2 \geq 40)$ emissions. For e.g. in the case of $^{14}$C emission from $^{248}$Cf, $T_{1/2}$= 1.31x10$^{41}$s; $^{14}$C emission from $^{249}$Cf, $T_{1/2}$ = 1.09x10$^{42}$s; $^{20}$O emission from $^{250}$Cf, $T_{1/2}$ = 2.81x10$^{55}$s; $^{10}$Be emission from $^{252}$Cf, $T_{1/2}$ = 4.87x10$^{78}$s etc., which are above the present experimental limit for measurements ($T_{1/2} \leq 10^{30} s$). Hence medium mass clusters C, O, Ne etc are not included in the Table 1. In the present calculation we rely on the $T_{1/2}$ that is of 2-3 units. But it is to be noted that the microscopic potential like Skyrme Energy density [38, 39] is also available that can change the $\log_{10}(T_{1/2})$ quite bit.

The nuclear proximity potential for oriented and deformed (with higher multipole deformation) nuclei are done following the prescription of Gupta and co-workers [35] with universal proximity potential given in eqns. 6 and 7. The Coulomb potential for the two deformed and oriented nuclei is computed using eqn. 15. In fission and cluster decay the fragments are strongly polarized due to nuclear force and accordingly their symmetry axes are aligned. In the present calculations we consider the pole to pole configuration for prolate daughter and spherical/prolate cluster. The proper inclusion of higher multipole deformations along with generalized orientation contributions may prove important in deciding the cluster decay paths of various clusters.

Table 1 represents the comparison of computed logarithm half life time for all the possible cluster emissions from $^{248-254}$Cf parents for the case of without deformation (a), with quadrupole deformation (b) and with quadrupole and hexadecapole ($\beta_2$ & $\beta_4$) deformations (c). The experimental deformation values $\beta_2$ taken from Ref [40] and for the cases in which there are only theoretical ones we have taken them from the tables of Moller et al [41]. It is obvious from the table that the half lives decrease with the inclusion of quadrupole deformation due to the fact is that it reduces the height and width of the barrier (increases the barrier penetrability). We would like to mention that the sign of hexadecapole deformation have no influence on half life time. In the case of oblate daughter and spherical/oblate cluster we have considered the equator-equator configuration because according to Gupta et al [36] the optimum orientation (lowest barrier) is equator-equator for oblate daughter and spherical/oblate cluster.

Fig.1 represents the comparison of computed alpha decay half life time (with and without including deformation) and experimental data. The experimental values are taken from [42]. It is also clear from the plot that the half life values (with including quadrupole deformation $\beta_2$) are in agreement with experimental data (for e.g. in $^{248}$Cf, $\log_{10} T_{Expt} = 7.46$, $\log_{10} T_{\beta_2} = 6.51$; in $^{250}$Cf $\log_{10} T_{Expt} = 8.0$, $\log_{10} T_{\beta_2} = 7.17$ etc.). It is to be noted that in the case of even-odd isotopes the half lives are found to be higher than that for even-even isotopes. Also in the case of $^{249}$Cf, $^{251}$Cf and $^{253}$Cf, as the experimental deformation values were unavailable; we have used the theoretical values from Ref [41]. This may be the reason why the potential without deformation gives better results than with deformation for these isotopes.

It is to be noted that the proximity theorem states that when surface thickness is comparable to the radius, this potential cannot be used. Thus the theorem limits the mass of the clusters to 16 or small. In Coulomb and Proximity Potential Model (CPPM), for the touching configuration and for the separated fragments, the potential energy barrier is taken as the sum of Coulomb potential, proximity potential and centrifugal potential. But for the pre-scission (overlap) region, simple power law interpolation eqn (10) is used instead of using the universal proximity potential given as

$$\phi(\xi) = -1.7817 + 0.9270\xi + 0.0143\xi^2 - 0.09\xi^3, \qquad \text{for } \xi \le 0 \tag{17}$$

This enables us to use our model for the half life calculations for clusters with mass $A_2 \le 16$, including alpha particle.

One of the interesting facts is that the $^{248\text{-}254}$Cf isotopes are instable against heavy cluster emissions and the corresponding daughter nuclei are doubly magic $^{208}$Pb (with Z=82, N =126) or neighbouring one. For e.g. different sulphur isotopes are probable for emissions from $^{248\text{-}252}$Cf isotopes, which stress the role of double magicity of the daughter nuclei $^{208}$Pb and various Ar clusters are possible in all the $^{248\text{-}254}$Cf isotopes which leads to the near doubly magic shell closures of daughter nuclei $^{206}$Hg (with $Z \approx 82$, N = 126). Various calcium isotopes are also probable for emission from these parents, which stress the role of doubly or near doubly magic $^{48}$Ca cluster and also the near double magicity of Pt daughter. i.e. the minimum half life time is obtained for $^{48}$Ca in $^{252}$Cf, $^{49}$Ca in $^{253}$Cf and $^{50}$Ca in $^{254}$Cf which indicates the role of neutron shell closures of the daughter nuclei $^{204}$Pt (N=126).

## 4. Conclusion

Below we summarize the main conclusions of the present study

1. It is found that the $^{248\text{-}254}$Cf parents are stable against light clusters (except alpha particle) and are instable against heavy cluster ($^{46}$Ar, $^{48,50}$Ca etc) emissions.

2. For the case of heavy cluster emissions the daughter nuclei are doubly magic $^{208}$Pb (with Z=82, N =126) or neighbouring one. Various calcium isotopes are also probable for emission from $^{248-254}$Cf parents, which stress the role of doubly or near doubly magic $^{48}$Ca cluster and also indicates the role of neutron shell closures of the daughter nuclei $^{204}$Pt (N=126).
3. The effects of quadrupole and hexadecapole deformations of both parent and fragments on half life times are studied using Coulomb and proximity potential for oriented and deformed nuclei as the interacting barrier.
4. The computed alpha decay half life values (with including quadrupole deformation $\beta_2$) are in agreement with experimental data.
5. For most of the cluster decays we have found that the half lives decrease with the inclusion of quadrupole deformation ($\beta_2$) due to the fact is that it reduces the height and width of the barrier (increases the barrier penetrability).
6. For the case of oblate daughter and spherical/oblate cluster we have considered the equator-equator orientation, because optimum orientation (lowest barrier) is equator-equator for oblate daughter and spherical/oblate cluster.

**References**

[1] A. Sandulescu, D. N. Poenaru and W. Greiner, Fiz. Elem. Chasitst. At. Yadra. **11** (1980) 1334 (Sov. J. Part. Nucl. **11** (1980) 528)

[2] H. J. Rose and G. A. Jones, Nature **307** (1984) 245

[3] A. H. Jaffey and A. Hirsch, unpublished data, quoted in: Vandenbosch and J. R. Huizenga, Nuclear Fission (Academic Press, New York, 1973)

[4] R. Bonetti, E. Fioretto, C. Migliorino, A. Paisnetti, F. Barranco, E. Vigezzi and R. A. Broglia, Phys. Lett B **241** (1990) 179

[5] A. Sandulescu, R. K. Gupta, F. Carstoiu, M. Horoi and W. Greiner, Int. J. Mod. Phys. E **1**

(1992) 379

[6] R. K. Gupta, M. Horoi, A. Sandulescu, W. Greiner and W. Scheid, J. Phys. G: Nucl. Part. Phys. **19** (1993) 2063

[7] G. A. Pik Pichak, Yad. Fiz. **44** (1986) 1421

[8] G. A. Pik Pichak, Sovt. J. Nucl. Phys. **44** (1986) 923

[9] Y. J. Shi andW. J. Swiatecki, Nucl. Phys. **A464** (1987) 205

[10] S. Kumar and R. K. Gupta, Phys. Rev. C **55** (1997) 218

[11] G. Shanmugam and B. Kamalaharan, Phys. Rev. C **38** (1988) 1377

[12] G. Shanmugam and B. Kamalaharan, Phys. Rev. C **41** (1990) 1184

[13] S. G. Thompson, K. Street, Jr., A. Ghiorso and G. T. Seaborg, Phys. Rev. **78** (1950) 298

[14] S. G. Thompson, K. Street, Jr., A. Ghiorso, and G. T. Seaborg, Phys. Rev. **80** (1950) 790

[15] H. Diamond, L. B. Magnusson, J. F. Mech, C. M. Stevens, A. M. Friedman, M. H. Studier, P. R. Fields, and J. R. Huizenga, Phys. Rev. 94 (1954) 1083

[16] F. Gonnenwein, A. Moller, M. Croni, M. Hesse, M. Wostheiarich, H. Faust, G. Fioni and S. Oberstedt, Nuovo Cimento A**110** (1997) 1089

[17] A. Moller, M. Croni, F. Gonnenwein and G. Petrov, in Proceedings of the International Conference on Large Amplitude Motion of Nuclei edited by C Giardina (Brolo, Italy, 1996)

[18] A. Sandulescu, A. Florescu, F. Carstoiu, A. V. Ramayya, J. H. Hamilton, J. K. Hwang, B. R. S. Babu and W. Greiner, Nuovo Cimento A **110** (1997) 1079

[19] A. Sandulescu, A. Florescu, F. Carstoiu, W. Greiner, J. H. Hamilton, A. V. Ramayya and B. R. S. Babu, Phys. Rev. C **54** (1996) 258

[20] R. K. Biju, Sabina Sahadevan, K. P. Santhosh and Antony Joseph, Pramana J. Phys. **70** (2008) 617

[21] Y. J. Shi and W. J. Swiatecki, Nucl. Phys. A **438** (1985) 450

[22] S. S. Malik and R. K. Gupta, Phys. Rev. C **39** (1989) 1992

[23] R. K. Gupta, S. Singh, R. K. Puri and W. Scheid, Phys. Rev. C **47** (1993) 561

[24] R. K. Gupta, S. Singh, R. K. Puri, A. Sandulescu, W. Greiner and W. Scheid, J. Phys G: Nucl. Part. Phys. **18** (1992)1533

[25] S. S. Malik, S. Singh, R. K. Puri, S. Kumar and R. K. Gupta, Pramana J. Phys. **32** (1989) 419

[26] J. Blocki, J. Randrup, W. J. Swiatecki and C. F. Tsang, Ann. Phys, NY **105** (1977) 427

[27] J. Blocki and W. J. Swiatecki, Ann. Phys., NY **132** (1981) 53

[28] I. Dutt and R. K. Puri, Phys. Rev. C **81**(2010) 044615

[29] I. Dutt and R. K. Puri, Phys. Rev. C **81**(2010) 047601

[30] I. Dutt and R. K. Puri, Phys. Rev. C **81** (2010) 064608

[31] I. Dutt and R. K. Puri, Phys. Rev. C **81**(2010) 064609

[32] K. P. Santhosh, Physica Scripta **81** (2010) 015203

[33] D. N. Poenaru, M. Ivascu, A. Sandulescu and W. Greiner, Phys. Rev. C **32** (1985) 572

[34] C. Y. Wong, Phys. Rev. Lett. **31** (1973) 766

[35] N. Malhotra and R. K. Gupta, Phys Rev. C **31** (1985) 1179

[36] R. K. Gupta, M. Balasubramaniam, R. Kumar, N. Singh, M. Manhas and W. Greiner, J. Phys. G: Nucl. Part. Phys. **31** (2005) 631

[37] G. Audi, A. H. Wapstra and C. Thivault, Nucl. Phys A **729** (2003) 337

[38] R. K. Puri, P. Chattopadhyay, and R. K. Gupta Phys. Rev. C **43** (1991)315

[39] R. K. Puri and R. K. Gupta, Phys. Rev. C **45** (1992) 1837

[40] http://www-nds.iaea.org/RIPL-2/

[41] P. Moller, J. R. Nix, W. D. Myers and W. J. Swiatecki, At. Data. Nucl. Data. Tables **59** (1995) 185

[42] G. Royer, J. Phys G. Nucl. Part. Phys. **26** (2000) 1149

Table 1. The comparison of calculated values of logarithm of half-life time for the case with out deformation (a), with qudrupole deformation $\beta_2$ (b) and with deformations $\beta_2$ & $\beta_4$ (c).

| Parent nuclei | Emitted cluster | Q value (MeV) | $\log_{10}(T_{1/2})$ without deform (a) | $\log_{10}(T_{1/2})$ with $\beta_2$ (b) | $\log_{10}(T_{1/2})$ with $\beta_2$ &$\beta_4$ (c) | Parent nuclei | Emitted cluster | Q value (MeV) | $\log_{10}(T_{1/2})$ without deform (a) | $\log_{10}(T_{1/2})$ with $\beta_2$ (b) | $\log_{10}(T_{1/2})$ with $\beta_2$ &$\beta_4$ (c) |
|---|---|---|---|---|---|---|---|---|---|---|---|
| $^{248}$Cf | $^{4}$He | 6.36 | 8.91 | 6.51 | 5.37 | $^{251}$Cf | $^{48}$Ar | 122.61 | 32.22 | 39.51 | 38.07 |
| | $^{40}$S | 111.85 | 27.29 | 30.16 | 27.94 | | $^{49}$Ca | 138.02 | 30.17 | 28.37 | 31.41 |
| | $^{44}$Ar | 124.20 | 31.96 | 29.89 | 30.44 | | $^{50}$Ca | 137.46 | 30.55 | 29.86 | 32.66 |
| | $^{46}$Ar | 124.32 | 30.94 | 30.64 | 30.91 | | $^{51}$Ca | 136.64 | 31.45 | 32.47 | 33.55 |
| | $^{48}$Ca | 138.07 | 31.35 | 30.21 | 29.66 | $^{252}$Cf | $^{4}$He | 6.22 | 9.61 | 7.17 | 6.54 |
| | $^{50}$Ca | 136.73 | 32.53 | 33.87 | 32.46 | | $^{42}$S | 107.97 | 32.87 | 35.06 | 35.62 |
| $^{249}$Cf | $^{4}$He | 6.23 | 9.24 | 7.17 | 6.17 | | $^{46}$Ar | 126.71 | 24.21 | 20.79 | 20.36 |
| | $^{40}$S | 110.20 | 30.31 | 32.32 | 30.30 | | $^{47}$Ar | 124.24 | 29.41 | 34.50 | 35.63 |
| | $^{41}$S | 110.08 | 29.86 | 31.68 | 30.56 | | $^{48}$Ar | 123.94 | 29.45 | 35.98 | 36.01 |
| | $^{42}$S | 109.39 | 30.68 | 33.26 | 32.90 | | $^{48}$Ca | 139.50 | 27.97 | 22.16 | 21.65 |
| | $^{44}$Ar | 124.29 | 31.62 | 27.49 | 26.98 | | $^{49}$Ca | 138.71 | 28.72 | 24.71 | 25.07 |
| | $^{45}$Ar | 124.15 | 31.31 | 36.48 | 34.76 | | $^{50}$Ca | 138.20 | 29.00 | 27.74 | 30.38 |
| | $^{46}$Ar | 124.72 | 29.65 | 29.46 | 30.50 | | $^{51}$Ca | 135.68 | 32.99 | 32.14 | 27.53 |
| | $^{47}$Ar | 122.99 | 32.41 | 38.78 | 36.43 | $^{253}$Cf | $^{4}$He | 6.13 | 10.08 | 7.64 | 7.14 |
| | $^{48}$Ca | 137.69 | 31.79 | 30.18 | 30.72 | | $^{46}$Ar | 125.25 | 26.66 | 24.85 | 22.38 |
| | $^{49}$Ca | 137.63 | 31.29 | 30.55 | 30.61 | | $^{47}$Ar | 126.17 | 24.10 | 33.37 | 35.10 |
| | $^{50}$Ca | 136.72 | 32.32 | 32.81 | 32.31 | | $^{48}$Ar | 124.80 | 26.33 | 36.26 | 38.37 |
| | $^{51}$Ca | 136.72 | 31.73 | 33.73 | 33.05 | | $^{48}$Ca | 138.03 | 30.36 | 26.14 | 24.64 |
| $^{250}$Cf | $^{4}$He | 6.13 | 10.11 | 7.69 | 6.79 | | $^{49}$Ca | 139.85 | 26.48 | 21.28 | 20.88 |
| | $^{40}$S | 108.76 | 32.61 | 33.97 | 32.28 | | $^{50}$Ca | 140.26 | 25.07 | 22.11 | 22.71 |
| | $^{42}$S | 110.13 | 29.11 | 24.24 | 25.83 | | $^{51}$Ca | 137.79 | 28.92 | 28.30 | 31.32 |
| | $^{44}$Ar | 124.39 | 31.23 | 26.34 | 25.39 | $^{254}$Cf | $^{4}$He | 5.93 | 11.18 | 8.83 | 8.37 |
| | $^{45}$Ar | 123.19 | 32.90 | 36.72 | 32.65 | | $^{46}$Ar | 124.16 | 29.77 | 25.97 | 25.81 |
| | $^{46}$Ar | 125.59 | 26.59 | 26.80 | 30.15 | | $^{47}$Ar | 123.48 | 30.52 | 37.30 | 39.10 |
| | $^{48}$Ca | 137.98 | 31.07 | 28.43 | 31.07 | | $^{48}$Ar | 125.46 | 24.86 | 34.97 | 38.09 |
| | $^{50}$Ca | 137.36 | 30.95 | 30.24 | 32.08 | | $^{48}$Ca | 136.92 | 32.14 | 25.97 | 25.79 |
| $^{251}$Cf | $^{4}$He | 6.18 | 9.85 | 7.40 | 6.80 | | $^{49}$Ca | 137.15 | 31.10 | 26.99 | 25.79 |
| | $^{42}$S | 108.96 | 31.16 | 33.59 | 33.65 | | $^{50}$Ca | 140.17 | 25.11 | 20.18 | 20.04 |
| | $^{45}$Ar | 124.81 | 29.78 | 34.66 | 35.03 | | $^{51}$Ca | 138.62 | 27.19 | 24.10 | 24.98 |
| | $^{46}$Ar | 126.15 | 25.40 | 22.44 | 22.54 | | $^{52}$Ca | 136.44 | 30.60 | 29.94 | 30.61 |

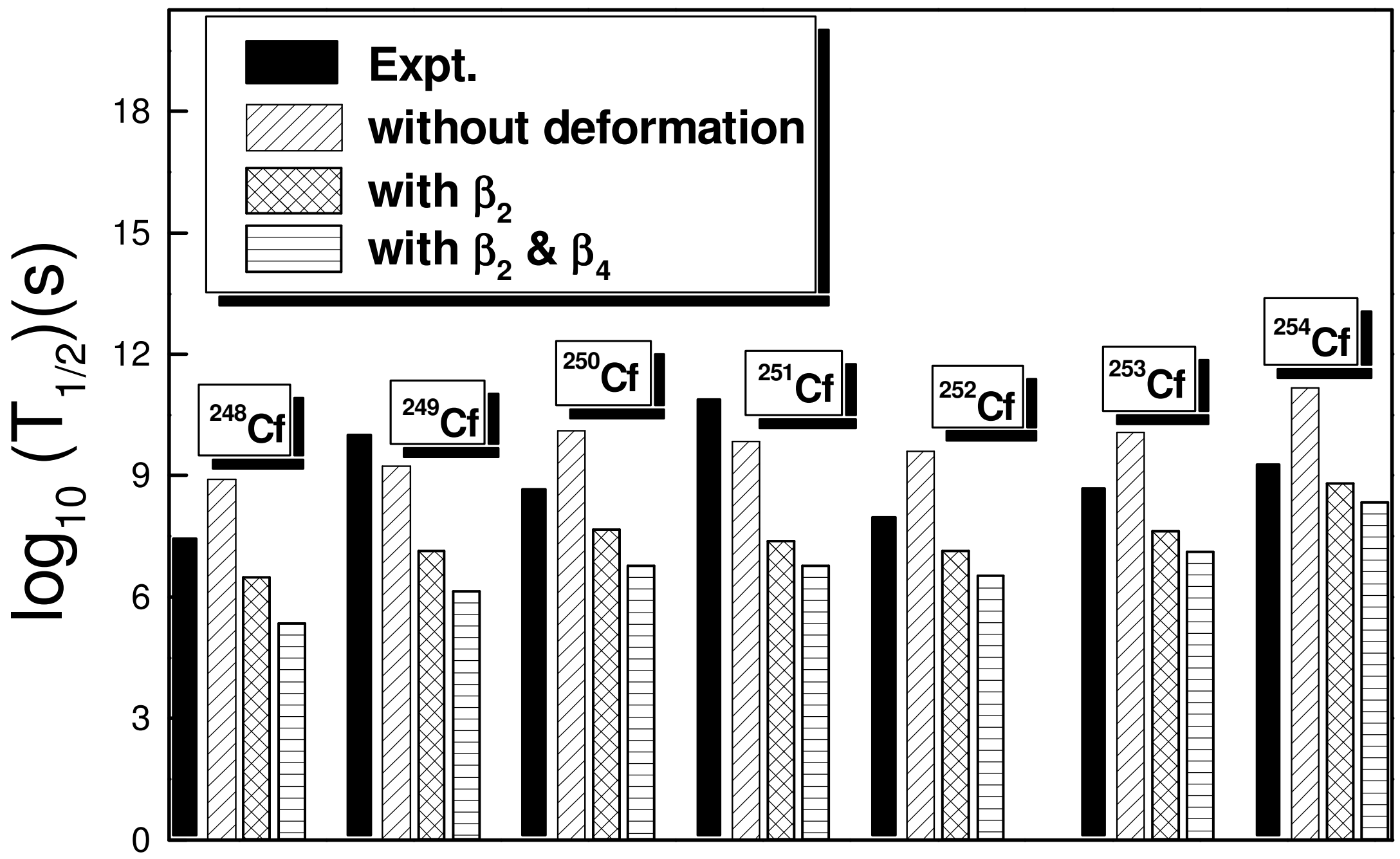

Fig .1. The comparison of computed alpha decay half life time with the experimental values